\documentclass[runningheads]{llncs}
\usepackage[T1]{fontenc}
\usepackage{graphicx}
\usepackage{tabularx}
\usepackage{enumitem}
\usepackage{amsmath}
\usepackage{amsfonts}

\begin{document}
\title{AI-based Multimodal Biometrics \\for Detecting Smartphone Distractions:   \\ Application to Online Learning}
\titlerunning{AI-based Multimodal Biometrics for Detecting Smartphone
Distractions}
%
\author{Alvaro Becerra\inst{1}\orcidID{0009-0003-7793-2682} \and
Roberto Daza\inst{1}\orcidID{0009-0005-2109-7782} \and
Ruth Cobos\inst{1}\orcidID{0000-0002-3411-3009} \and
Aythami Morales\inst{1}\orcidID{0000-0002-7268-4785} \and
Mutlu Cukurova\inst{2}\orcidID{0000-0001-5843-4854} \and
Julian Fierrez\inst{1}\orcidID{0000-0001-5843-4854}}
\authorrunning{Becerra et al.}
%
\institute{Universidad Autónoma de Madrid, Spain \email{\{alvaro.becerra, roberto.daza,ruth.cobos,aythami.morales,julian.fierrez\}@uam.es} \and University College London, United Kingdom \email{m.cukurova@ucl.ac.uk}}
\maketitle              
\begin{abstract}
    This work investigates the use of multimodal biometrics to detect distractions caused by smartphone use during tasks that require sustained attention, with a focus on computer-based online learning. Although the methods are applicable to various domains, such as autonomous driving, we focus on the challenges learners face in maintaining engagement amid internal (e.g. motivation), system-related (e.g., course design) and contextual (e.g., smartphone use) factors. Traditional learning platforms often lack detailed behavioral data, but Multimodal Learning Analytics (MMLA) and biosensors provide new insights into learner attention. We propose an AI-based approach that leverages physiological signals and head pose data to detect phone use. Our results show that single biometric signals, such as brain waves or heart rate, offer limited accuracy, while head pose alone achieves 87\%. A multimodal model that combines all signals reaches accuracy 91\%, highlighting the benefits of integration. We conclude by discussing the implications and limitations of deploying these models for real-time support in online learning environments.

\end{abstract}
\section{Introduction}
In recent years, the popularity of online courses has increased exponentially, especially during and after the COVID-19 pandemic \cite{sharin2021learning}. This shift to online education has provided learners with greater flexibility and access to resources. However, it has also introduced several unique challenges. One of the primary difficulties facing learners in these online environments is the lack of direct interaction with instructors \cite{iraj2020understanding}. The absence of this personal connection in virtual classrooms can negatively impact the motivation of learners, contributing to a sense of isolation and, in many cases, resulting in higher dropout rates \cite{topali2019exploring}.

In addition, online learners are particularly vulnerable to distractions that significantly impact their focus and overall learning outcomes. As demonstrated by research, momentary interruptions, even as brief as a few seconds, can significantly increase error rates by disrupting the cognitive processes required to maintain task sequence and continuity \cite{altmann2014momentary}. Distractions during online learning sessions are often caused by both external and internal factors, such as environmental disturbances and low levels of engagement, which can reduce concentration and hinder the absorption of valuable information \cite{betto2023distraction}.

These distractions may include constant phone notifications, emails, and other external interruptions that are less prevalent in traditional classroom settings. Such distractions are particularly pervasive in online learning environments, where the boundary between academic and non-academic tasks often blurs. Multitasking, such as responding to emails or participating in social networks during learning sessions, significantly reduces cognitive efficiency and impairs information retention, leading to measurable declines in academic performance \cite{winter2010effective,blasiman2018distracted}. In addition, passive distractions, such as background noise or household chores, can consume attentional resources and further detract from the learning process \cite{blasiman2018distracted}. Furthermore, many factors that emanate from the physical learning space, such as the presence of others, the light, or the spatial comfort, also affect online learning \cite{ciordas2021mobile}. As a result, maintaining concentration and staying on task becomes a major challenge for learners in online education, further complicating the learning process and affecting academic performance.

Understanding, detecting, and predicting when learners become distracted during online learning could be very beneficial, as it would allow to provide personalized feedback to guide them in reviewing content when distractions occur. Several articles have shown that giving general feedback to learners, using only data from their interactions with an online course, improves their performance \cite{becerra2024generative,cobos2023selfregulated,topali2024codesign}. However, relying solely on interactions data has several limitations, so biometric data, biosensors \cite{2020_CDS_HCIsmart_Acien}, and new online learning platforms \cite{daza2023edbb,2015_ITiCSE_Kesytroke_Aythami}, which incorporate biometrics and behavioral analysis, have emerged, offering a deeper understanding of learners' cognitive, emotional states, and overall behavior. This technology enables the detection of user distraction events more effectively.

Physiological and biometric signals have proven to be highly valuable in understanding the state of the learner, such as cognitive load \cite{daza2024deepface}, blinking of the eyes \cite{daza2024mebal2}, facial expression \cite{daza2025smartevr}, and pose of the learner \cite{goldberg2021attentive}, among others. These signals are especially valuable in online learning platforms, and when integrated with data from learners' interactions in an online course, they offer significant potential for enhancing predictive models.
Ultimately, when these technologies are integrated with the MMLA tools, a more comprehensive insight into the behavior of the learner can be obtained \cite{Mangaroska2021}.

In this article, our aim is to better understand distractions, particularly distractions from the mobile phone. The main contributions of this paper are:

\begin{itemize}
\item The development of various unimodal machine learning approaches for predicting mobile phone usage, using global features within a 40-second window. Each model uses a single signal, such as heart rate, attention, meditation, EEG waves, or head pose. The results are analyzed to determine which signal provides the highest prediction accuracy.
\item The proposal of two new multimodal systems for the detection of phone usage, one of which achieves the accuracy 91\%, surpassing unimodal systems.
\end{itemize}

\section{Related Works}\label{SEC:related_works}

AI-based approaches have become essential tools in MMLA, enabling the detection of complex student behaviors. By integrating multimodal data, machine learning algorithms can uncover hidden patterns that can lead to the identification of behaviors such as moments of distraction.

\subsection{Multimodal Learning Analytics}

Advances in Multimodal Learning Analytics (MMLA) have enabled a deeper understanding of learner behavior by combining data from multiple biosensors and interaction sources. These approaches integrate biometric signals, behavioral tracking, and machine learning models to improve learning outcome predictions, detect attention levels, and identify different events \cite{giannakos2019multimodal}.

Biometric and multimodal data collected from various sensors \cite{2023_SNCS_Human-Centric_Pena} have proven to be highly valuable in gaining deeper insights into learners' interactions and success. \cite{spikol2018supervised} demonstrated that the integration of multimodal data streams, such as facial tracking, hand movements, and sensor inputs in real-time, enables the application of supervised machine learning techniques to predict learner outcomes in project-based learning environments. Recent research, for example, \cite{yan2025complexity} showed how data from multiple sensors, such as position trackers, microphones, and heart rate monitors, can be combined to identify clear patterns of learner behavior while students interact and collaborate during learning tasks. Similarly, in \cite{becerra2025mosaic}, data from webcams, microphones, eye trackers, and wearable sensors are used to generate personalized data-based feedback on students’ oral presentation skills.

The variables of the students studied the most in MMLA include performance outcomes, attention levels, and behavior detection. For example, in \cite{navarro2024vaad}, eye tracking data, specifically fixations and saccades, are analyzed to assess the visual attention of learners and predict the tasks they are engaged in. Other studies target specific behavioral events, for example, \cite{kaddoura2022towards} used videos from front and back webcams, combined with speech analysis, to detect cheating behaviors during online exams using a deep convolutional neural network. Recent studies have also examined the development of MMLA dashboards to visualize and synchronize multimodal data, as well as their integration with Generative AI technologies \cite{becerra2025aicofe,becerra2025m2lads}.

\subsection{Detection of Distraction in Digital Learning Environments}
Recent studies have focused on detecting distractions in online learning through the use of biosensors and interaction data \cite{2020_CDS_HCIsmart_Acien}. In \cite{rodriguez2024application}, mouse dynamics \cite{2022_PR_BeCAPTCHA-Mouse_Acien}, including clicks, movements, and idle times, are analyzed to detect moments of distraction in students who interact with online learning platforms. Using techniques such as the DBSCAN clustering algorithm and PrefixSpan sequential pattern mining, the study identifies frequent interaction patterns that precede distraction events, such as browsing social networks instead of focusing on educational material. The results demonstrate an effective segmentation of the students into distinct groups. \cite{betto2023distraction} utilized machine learning to detect student distraction during online learning sessions by extracting facial feature points and postural data from webcam recordings. Their method incorporated OpenFace and GAST-Net to track eye movements, head posture, and skeletal coordinates, achieving greater 90\% recall in binary classification tasks. In this classification, the states of the students were classified into ``distracted'' and ``concentration'' states. The labeling of distraction events was determined using skin conductance responses (SCRs) obtained from wristband sensors, where distraction was defined as periods in which SCR occurred four or more times per minute. This physiological approach enabled an objective labeling mechanism that eliminates the dependency on self-reported surveys.

Other studies have focused on predicting learner attention levels \cite{daza2024deepface,daza2021alebk,goldberg2021attentive,peng2020predicting}, as low attention levels can be associated with distractions. For example, \cite{daza2024deepface} proposed a method for estimating attention levels (cognitive load) through an ensemble of facial analysis techniques applied to webcam videos using machine learning. This model used features such as facial expressions, blinking of the eyes, and posture to predict whether the learner's attention was high or low.

Expanding on distraction detection in immersive environments, \cite{asish2022detecting} focused on detecting when learners became distracted while using Virtual Reality for education. They used eye gaze tracking to measure attention shifts in response to external distractions, such as social media notifications, mobile ringtones, and background conversations. Their approach used a CNN-LSTM deep learning model, achieving 89.8\% accuracy in classifying distraction into three levels (low, medium, high). Furthermore, they demonstrated that Random Forest models achieved even higher accuracy (98.88\%) when incorporating gaze angles and head-tracking data.

\subsection{Detection of Phone Usage Events}
The detection of phone usage events has been studied, primarily in the context of driver distraction detection. An example is the work by Artan et al. \cite{artan2014driver}, which proposes a computer vision-based method to detect the usage of driver cell phones using a near-infrared (NIR) camera system directed at the front windshield of the vehicle. Their approach consists of two main stages: (1) face localization using a deformable parts model (DPM) and (2) phone usage classification through image classification techniques, such as Fisher vectors, combined with a linear Support Vector Machine (SVM). Their system was tested on a dataset of 1,500 front-view vehicle images, achieving 86.19\% accuracy with the best model.

Another relevant study is the work of Seshadri et al. \cite{seshadri2015driver}, which presents a computer vision-based system to detect driver phone usage using face-view videos. Their method employs the Supervised Descent Method to track 49 facial landmarks and extract regions of interest around the driver's ear and hand. These regions are then classified using the Histogram of Oriented Gradients features combined with the Real AdaBoost, SVM, and Random Forest classifiers. Their system was tested on 13,023 video frames, achieving 93.86\% accuracy, outperforming previous approaches.

In addition to computer vision-based systems, Mendez et al. \cite{mendez2019mobile}, proposed a mobile application that detects phone usage while driving by analyzing accelerometer and GPS data collected directly from the smartphone. Although mobile phone usage detection has been extensively explored in driving safety research, its application in online learning environments has received little attention. One of the few existing studies is that of Becerra et al. \cite{becerra2024biometrics}. Their work focuses on detecting phone usage events by analyzing head pose deviations captured from webcam images during online learning. Using a semi-supervised approach, the system flags potential phone usage events for further human review and found that head posture changes significantly when learners engage with their phones, deviating from typical online learning behavior. 

These findings highlight the growing need for automated phone usage detection in online learning, as mobile phones are a significant source of distraction in digital education settings \cite{wang2022comprehensively}.

\section{MultiModal Data Description}\label{SEC:multimodal_data}
For this study, the IMPROVE database was used \cite{daza2024improve}, which provides multimodal and biometric data from learners participating in an online course. This database was selected because of its focus on studying the effects of mobile phone usage on learner performance in online courses.

In this article, out of the 120 IMPROVE learners, we focused on the 40 who use mobile phones and the 40 who do not (phone removed), since the remaining 40 learners belong to a third different experimental condition. Furthermore, of those 80 learners, those whose EEG band data showed more than 5 minutes of data loss were also excluded (7 from mobile phone users and 7 from nonusers). The 5-minute threshold was chosen because the sessions lasted between 25 and 30 minutes, which means that such a loss would represent approximately 15\% to 20\% of the total duration of the session. So, in total, we used 33 learners who used their phones during the learning session (LS) and 33 learners who did not have access to their phones. The learners who used their phones were required to keep them visible during the LS and were instructed to use them upon receiving notifications from messaging or email applications. Each learner received two messages from the researchers, and the periods during which they responded were labeled as phone events. The data used in this study from the IMPROVE database include:

\begin{itemize}
    \item \textbf{Head pose (\(^{\circ}\))
}: Roll, yaw, and pitch angles estimated per frame using a head pose detector applied to video captured by a front-facing webcam.
    \item \textbf{Heart rate (bpm)}: Heart rate data were collected at a frequency of 1Hz using a Huawei Watch 2 pulsometer.
    \item \textbf{Attention (0-100), meditation (0-100), and brain waves (dB)}: Signals were recorded using a NeuroSky EEG band at a frequency of 1Hz, which provided five frequency bands: $\delta$~($<4$ Hz), $\theta$~($4$-$8$ Hz), $\alpha$~($8$-$13$ Hz), $\beta$~($13$-$30$ Hz), and $\gamma$~($>30$ Hz). Using the official NeuroSky SDK, attention (focus level) and meditation (mental calmness) were also obtained.
    \item \textbf{Events data}: Information related to the activities the students performed during the LS, including using the phone, was collected.
\end{itemize}

Other IMPROVE data, such as keystroke activity \cite{2022_TBIOM_TypeNet_Acien,2025_PR_KVC} or eye tracking information, were not included in this study, as students using a mobile phone do not type on the computer or focus on the screen. Furthermore, thanks to the M2LADS system \cite{becerra2023m2lads,becerra2025m2lads}, all data were processed, synchronized, and labeled according to the activity the learner was doing. These activities include watching videos, reading content, solving assignments, or using a phone.

\section{Methods}\label{SEC:methods}

\subsection{Phone Event Detection Models}
We propose an AI-based approach to detect phone usage in learners using different physiological signals (attention, meditation, heart rate, and brain waves: alpha, beta, gamma, delta, and theta) and head pose data (roll, yaw, and pitch angles). This approach aims to identify whether a learner used the phone in the last 20 seconds (\( W_{\text{phone}} \)) of a 40-second window (\( W_{\text{all}} \)). A 20-second window was chosen because shorter windows may miss relevant phone usage patterns, while approximately 25\% of phone usage events lasted less than 20 seconds. Therefore, using longer windows would require including post-usage data, which may dilute the physiological signals directly related to the phone interaction. A binary classification is proposed to distinguish between phone use and non-use for each \( W_{\text{all}} \). To obtain the datasets, the following protocol was followed: 
\begin{enumerate}
    \item For learners who used the phone, the \( W_{\text{all}} \) windows contained 20 seconds before phone use (\( W_{\text{pre}} \)) and the first 20 seconds (\( W_{\text{phone}} \)) during phone use.
 In cases where the phone use lasted less than 20 seconds, \( W_{\text{phone}} \) included post-usage data. The pre-event window \( W_{\text{pre}} \) was considered useful for detecting changes in physiological signals triggered by phone usage.
  \item For learners who did not use the phone,  \( W_{\text{all}} \) were selected during the same activities as those who used the phone, specifically during the second video and while reading code. In this case, \( W_{\text{all}} \) consists of two 20-second segments (\( W_{\text{nophone}} \)) representing periods of no phone use.

  \end{enumerate}
 In total, we obtained 66 samples for mobile phone usage and 66 samples for no mobile phone usage. To predict events of mobile phone usage, we proposed an approach based on global characteristics, which have demonstrated effectiveness in other classification tasks \cite{daza2024deepface,fierrez2005line,martinez14mobileSignRobustPerf}.

\begin{table}[t]
    \centering
    \caption{Description of the global features where velocity is the first derivative of the signal, acceleration is the second, and jerk is the third. The 33 global features are calculated for all input signals $n \in \{1, 2, \ldots, S\}$, with $S=11$ (see Fig.~\ref{fig:diagrama_modelos_individuales} top).}
    \label{TB:caracteristicas}
    \scriptsize
    \begin{tabular}{|c|l|c|l|}
        \hline
        \textbf{\#} & \textbf{Feature Description} & \textbf{\#} & \textbf{Feature Description} \\
        \hline
        $g_n^1$ & Total positive velocity $\sum(v > 0)$ & $g_n^{18}$ & Maximum jerk $j_{\text{max}}$ \\
        $g_n^2$ & Total negative velocity $\sum(v < 0)$ & $g_n^{19}$ & RMS of jerk $j_{\text{RMS}}$ \\
        $g_n^3$ & (Location of the first maximum) & $g_n^{20}$ & Location of the maximum absolute jerk \\
        $g_n^4$ & (Location of the second maximum) & $g_n^{21}$ & Location of the maximum jerk \\
        $g_n^5$ & (Location of the third maximum) & $g_n^{22}$ & Number of sign changes in velocity \\
        $g_n^6$ & (Average velocity $v$) / $|v|_{\text{max}}$ & $g_n^{23}$ & $\sum(v > 0) |v| / \sum(v < 0) |v|$ \\
        $g_n^7$ & (Average velocity $v$) / $v_{\text{max}}$ & $g_n^{24}$ & $\sum(v > 0) / \sum(v < 0)$ \\
        $g_n^8$ & (RMS of velocity $v_{\text{RMS}}$) / $|v|_{\text{max}}$ & $g_n^{25}$ & $x_{\text{max}} - x_{\text{min}}$ \\
        $g_n^9$ & (Mean of the segment $\overline{x}$) & $g_n^{26}$ & $\overline{v} / (x_{\text{max}} - x_{\text{min}})$ \\
        $g_n^{10}$ & (Standard deviation of the segment $\sigma$) & $g_n^{27}$ & (Local maxima in segment x) \\
        $g_n^{11}$ & (RMS of acceleration $a_{\text{RMS}}$) / $|a|_{\text{max}}$ & $g_n^{28}$ & (Average acceleration $\overline{|a|}$) \\
        $g_n^{12}$ & Median of the segment & $g_n^{29}$ & Maximum of the segment \\
        $g_n^{13}$ & Standard deviation of velocity $\sigma_v$ & $g_n^{30}$ & Minimum of the segment \\
        $g_n^{14}$ & Standard deviation of acceleration $\sigma_a$ & $g_n^{31}$ & Skewness of x \\
        $g_n^{15}$ & Jerk $|j|$ & $g_n^{32}$ & Kurtosis of x \\
        $g_n^{16}$ & Average jerk & $g_n^{33}$ & Learner's gender \\
        $g_n^{17}$ & Absolute value of the maximum jerk $|j|_{\text{max}}$ & & \\
        \hline
    \end{tabular}
    \normalsize
\end{table}

Each \( W_{\text{all}} \) window was divided into two segments of 20 seconds. For each 20-second segment and each input signal $n \in \{1, 2, \ldots, S\}$ we calculate a global vector $\textbf{g}_n$ that includes 33 global features, as indicated in Table \ref{TB:caracteristicas}. The features of the two windows (either $\{\textbf{g}^{\text{pre}}_n,\textbf{g}^{\text{phone}}_n\}$ or $\{\textbf{g}^{\text{nophone}}_n,\textbf{g}^{\text{nophone}}_n\}$) are combined into a single vector \cite{2018_INFFUS_MCSreview1_Fierrez}, $\textrm{\textbf{f}}^{\mathrm{g}}_{\text{cmb}} \in \mathbb{R}^{65}$ (the learner's gender was only included once) , which is normalized to $\tilde{\textbf{f}}^{\mathrm{g}}_{\mathrm{cmb}}$ using the z-score technique \cite{Fierrez-Aguilar2005_ScoreNormalization}.

\begin{figure}[ht]
    \centering
    \includegraphics[width=\linewidth]{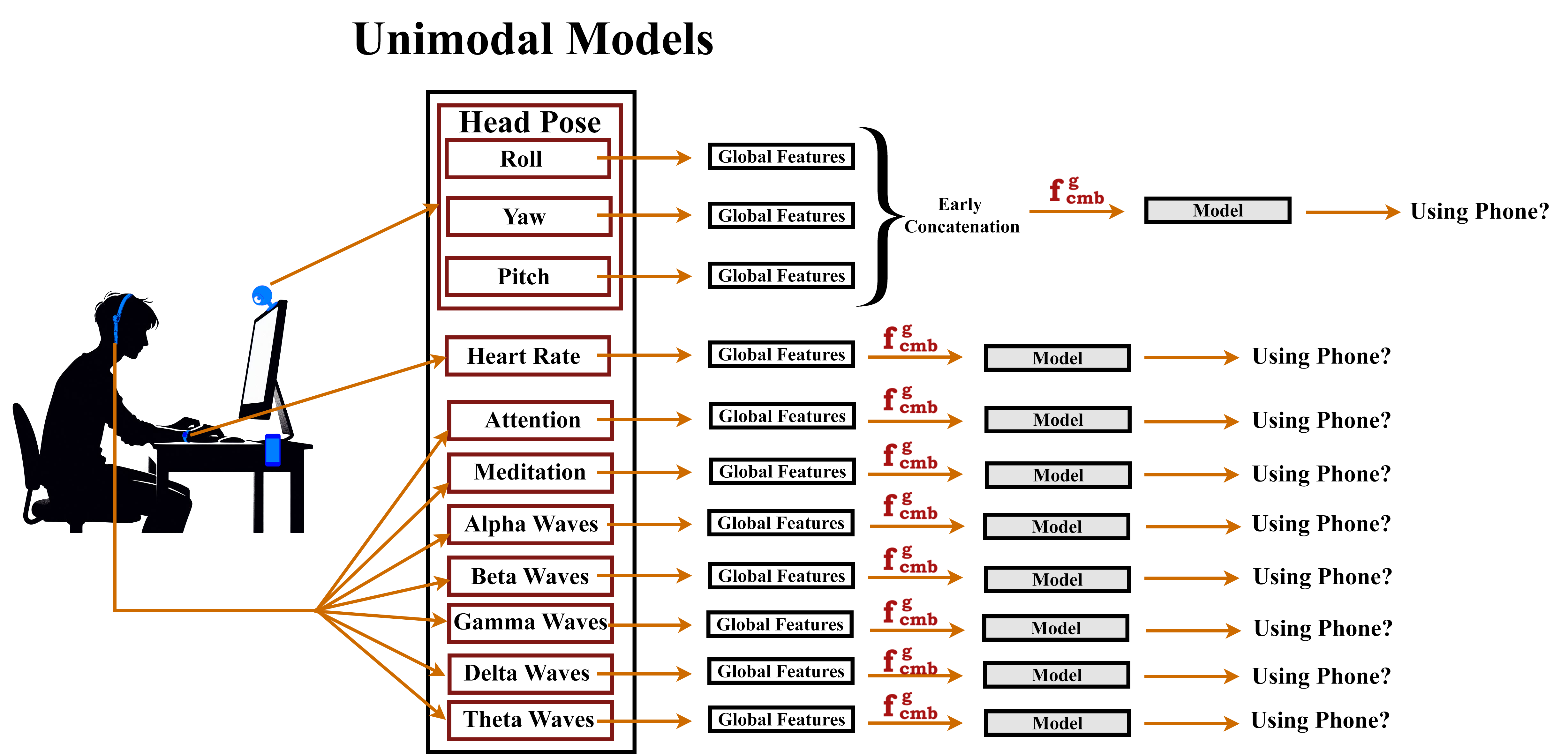}
    \caption{Outline of the unimodal models. For the EEG signals and heart rate we developed unimodal models using global features (Table \ref{TB:caracteristicas}). For the  model that use head pose angles, global features are extracted from roll, yaw and pitch and they are concatenated into a single vector. We also developed two multimodal models: one combining all EEG signals and Heart Rate and another combining all multimodal data}
    \label{fig:diagrama_modelos_individuales}
\end{figure}

As shown in Figure \ref{fig:diagrama_modelos_individuales}, we developed one model for each of the input biometric signals. For models using EEG signals and heart rate, two approaches were tested to obtain training data due to the noise present in the signals: using the raw data provided by the devices and using the same data but applying a smoothing window. If the smoothing window was size \(N\), the value at time \(t\) was calculated as the average of the previous values \(N\), including the one at \(t\). In this way, different sizes of smoothing windows were compared to determine their impact on signal quality.

Finally, we developed two additional models: one using data from all EEG signals and heart rate, and the other incorporating all multimodal data. For these models, an early concatenation of the individual global features was used to obtain the training vector. Different smoothing window sizes were also compared.

For each model developed, we evaluated two classification algorithms: Random Forest (RF) and Support Vector Machine (SVM), both of which have been widely used and validated in the literature for physiological signal classification tasks \cite{daza2024deepface,tatar2023biometric}. SVMs were trained using linear and Gaussian kernels with a regularization hyperparameter $C = 1$. RFs were trained with 250 trees after testing between 100 and 500 trees and observing no further improvement in performance beyond 250 trees. We applied two techniques for feature selection and dimensionality reduction: SelectKBest and PCA. Thus, we tested RF, SVM with a linear kernel, and SVM with a Gaussian kernel using all features, a subset of features selected with SelectKBest, and PCA with the number of components required to explain 95\% of the variance. Furthermore, algorithms were tested with smoothed and unsmoothed data using window sizes (\( W_{\text{smooth}} \)) of 5, 10, 15, 20, 25, and 30 seconds.

As the data for each variable consisted of 132 periods - 66 with mobile phone use and 66 without - Leave-One-Out cross-validation was selected at the participant level to maximize the amount of data used for modeling training. In each fold, data from a single learner (comprising two samples) was used as the test set, while the remaining 65 learners (130 samples) were used for training. This strategy ensures that no data from the test subject appear in the training set, avoiding data leakage and providing a more realistic evaluation of model generalization.

\begin{table}[h]
    \centering
    \caption{Accuracy results and the best models for phone usage detection}
    \label{TB:model_results}
    \begin{tabularx}{\textwidth}{|X|X|c|}
        \hline
        \textbf{Biometric Variable} & \textbf{Best Model} & \textbf{Accuracy (\%)} \\
        \hline \hline
        Attention     & SVM (linear kernel) with PCA and smoothing window of size 20 & 64\% \\
        \hline
        Heart Rate (HR)         & SVM (Gaussian kernel) with a subset of 40 features and smoothing window of size 5  & 70\% \\
        \hline
        Meditation   &  Random Forest with PCA and smoothing window of size 20  & 61\% \\
        \hline
        Alpha         & SVM (Gaussian kernel) with all features and smoothing window of size 20 & 64\% \\
        \hline
        Beta & Random Forest with all features and without smoothing window   & 68\% \\
        \hline
        Gamma         & Random Forest with all features and smoothing window of size 20 & 70\% \\
        \hline
        Theta         & SVM (linear kernel) with all features and smoothing window of size 30 & 61\% \\
        \hline
        Delta & SVM (linear kernel) with PCA and smoothing window of size 25   & 64\% \\
        \hline
        All EEG signals and heart rate (EEG+HR) & Random Forest with all features and smoothing window of size 15 & 76\% \\
        \hline
        Head Pose (HP) & Random Forest with a subset of 120 features & 87\% \\
        \hline
        All EEG signals, heart rate, and head pose (EEG+HR+HP) & Random Forest with a subset of 250 features and smoothing window of size 30 & 91\% \\
        \hline
    \end{tabularx}
\end{table}

\section{Results and Discussion}\label{SEC:results}

\subsection{Performance for Phone Detection Models}
Table \ref{TB:model_results} displays the results for the different models developed. Accuracy was selected as the primary evaluation metric due to the balanced nature of the dataset and the objective of binary classification to detect mobile phone usage. The findings suggest that the use of a single physiological signal leads to poorer model performance. In particular, models based on attention, meditation, alpha waves, theta waves, and delta waves exhibit the lowest accuracy. In contrast, models that incorporate heart rate, beta waves, and gamma waves perform slightly better, achieving around 70\% accuracy. These results indicate that mobile phone usage has a stronger impact on certain physiological signals than others.

However, individual signals were not highly effective in detecting phone usage. By combining all EEG signals with heart rate data (EEG + HR), the model performance improved significantly by 8.57\% (McNemar’s test, $\chi^2 = 24.45, p < 0.001$), reaching an accuracy of 76\%. This highlights that integrating multiple physiological signals offers a more comprehensive and reliable approach to detecting phone usage.

\begin{figure}[ht]
    \centering
    \includegraphics[width=\linewidth]{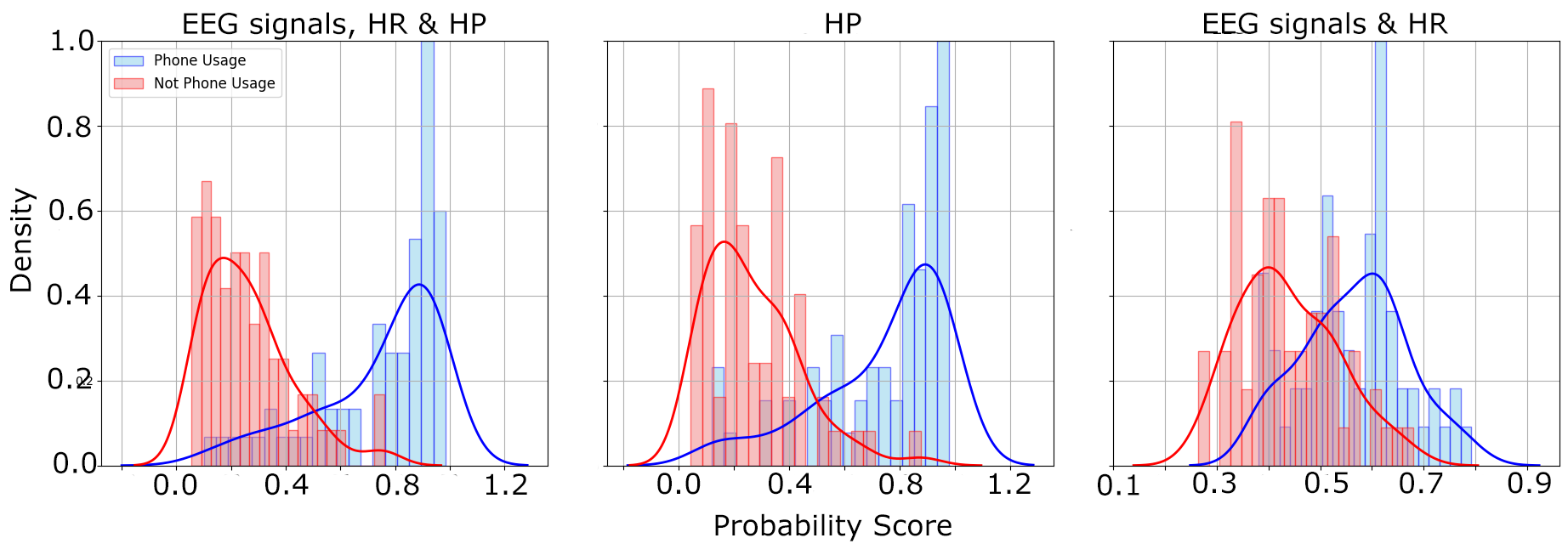}
    \caption{Probability density distributions of the probability scores obtained with the 3 best models indicated in Table \ref{TB:model_results}}
    \label{fig:densidad}
\end{figure}

The head pose model (HP) achieved an accuracy of 87\%, improved on the best unimodal models by 24.29\% (McNemar’s test, $\chi^2 = 31.58, p < 0.001$) and outperformed the model combining all EEG signals and heart rate by 14.47\% (McNemar’s test, $\chi^2 = 21.02, p < 0.001$). This result highlights the importance of head pose data, which often reveal notable changes in posture when interacting with a phone, compared to typical online learning behavior on a computer, making it particularly effective in detecting phone usage \cite{becerra2024biometrics}.

However, the most effective model was the one that combined all signals (EEG + HR + HP), achieving an accuracy of 91\%. This model improved on the head pose model by 4.60\% (McNemar’s test, $\chi^2 = 7.61, p = 0.0058$), further confirming that integrating signals from multiple sources leads to better predictive outcomes.

It should be noted that most models achieve the best results by smoothing the signals captured from the biosensors. This smoothing helps reduce the noise in the data and utilizes information from a larger set of data, making global features more relevant \cite{daza2024deepface}.

Figure \ref{fig:densidad} illustrates the probability density distributions of the probability scores generated by the three best models: one that combines all EEG signals, heart rate, and head pose (EEG+HR+HP); another using only head pose data (HP); and the third combining all EEG signals with heart rate (EEG+HR). To assess the separation between the two classes, Phone usage and Not Phone usage, we calculated the Kullback-Leibler (KL) divergence. The results showed a KL divergence of $0.22$ for the EEG+HR+HP model, a divergence of $0.24$ for the HP model and a divergence of $0.04$ for the  EEG+HR model. These results indicated that the head pose-only model achieved promising class separation performance, even when compared to a multimodal model. 

\begin{figure}[ht]
    \centering
    \includegraphics[width=0.6\linewidth]{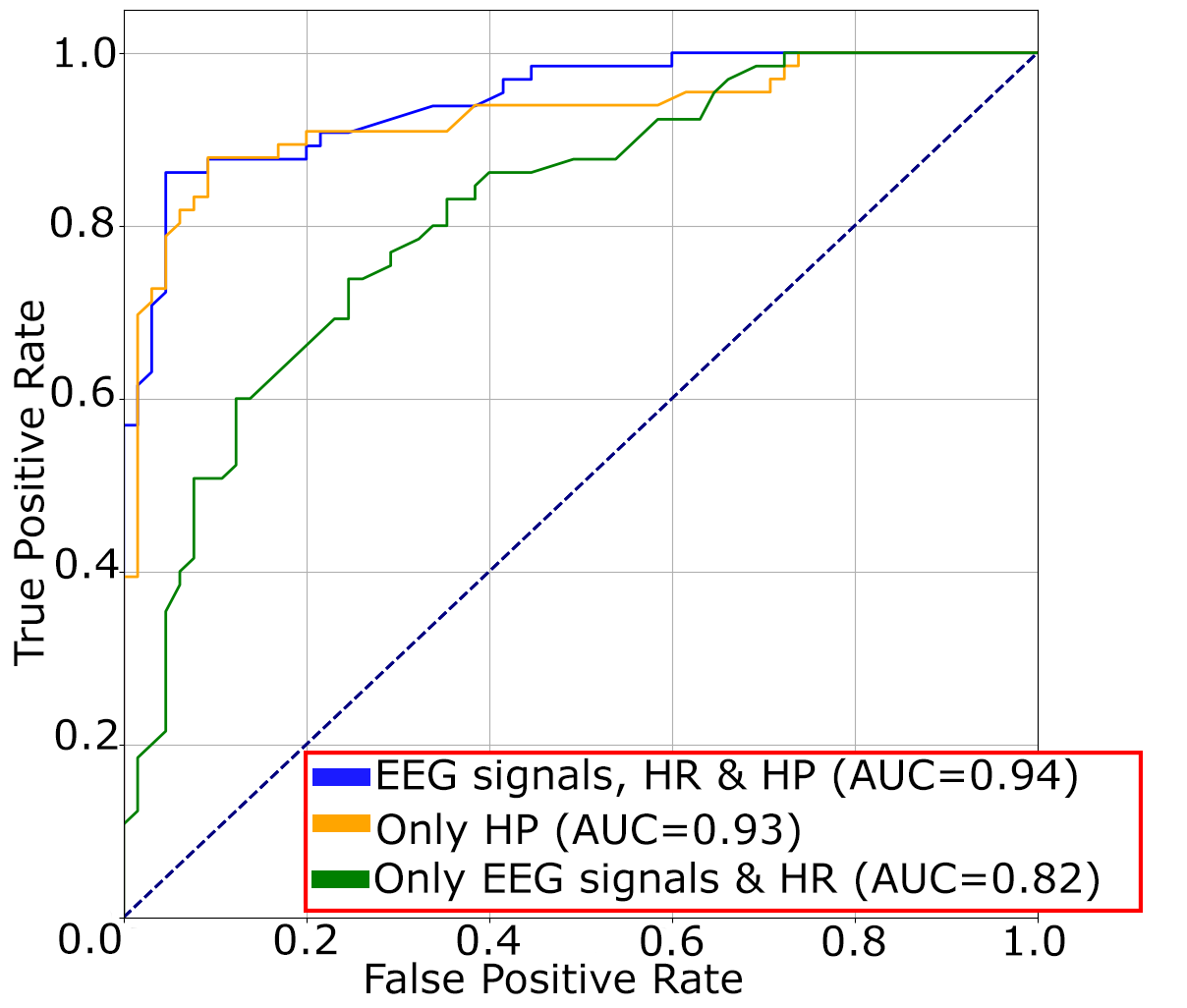}
    \caption{ROC obtained for the
most accurate models indicated in Table \ref{TB:model_results}}
    \label{fig:roc}
\end{figure}

However, the fact that the distributions are somewhat more separated does not necessarily mean that the performance is significantly better or worse (in fact, in terms of accuracy, combining all multimodal data achieves higher performance). In Figure \ref{fig:roc}, the ROC curves are displayed for the three models, showing that depending on the threshold, combining all multimodal data or using only the head pose may be the best option. In terms of Area Under the Curve (AUC), combining all multimodal data is slightly better than using only the head pose, but both models performed well overall.

\subsection{Practical Implications and Limitations}
Although the multimodal model integrating EEG signals, heart rate, and head pose achieved the highest accuracy (91\%), it is important to note that the head pose-based model also demonstrated strong performance. This suggests that movement patterns alone provide meaningful insight into phone usage detection.

A key advantage of the head pose model is its practicality for real-world applications. Unlike EEG or heart rate monitoring, which may require specialized and intrusive biosensors \cite{2020_MediComp_HRcomparison_JHO}, estimation of head posture can be performed using standard webcams, making it a more accessible and scalable solution. This opens the door to integration into on-line learning platforms that already track head movements or engagement levels. By incorporating head pose analysis, these platforms could provide real-time feedback to learners, helping them stay focused and minimizing phone-induced distractions.

Given the application limitations of the model that combines all biometric signals, future work could explore estimating these variables through the webcam. If these estimations are reliable, it would be possible to integrate a proxy version of the multimodal model without requiring additional wearable sensors. This approach would not only maintain ease of implementation and user comfort, but would also enhance the generalizability of the models by enabling data collection through more accessible and cost-effective methods. By eliminating the need for expensive biosensors, this solution could facilitate the widespread adoption of distraction detection systems in online learning environments, making advanced analytics available to a broader range of users.

Although the number of samples in the IMPROVE dataset is not exceptionally large (132 instances), it is still considered a relatively substantial dataset within the MMLA field. This study has shown that it is feasible to develop accurate models to detect smartphone usage using multimodal data. As a result, this work provides an empirical basis for future research efforts to scale and refine these models. Subsequent studies may focus on increasing the number of participants to improve the generalizability of the model, potentially relying solely on more accessible modalities such as the estimation of the head pose.

\section{Conclusion and Future Work}\label{SEC:conclusion}
This article explores the potential of MMLA to enhance the detection of phone usage among learners by using various biometric variables. Several AI-based models were proposed and evaluated to predict when learners use their phones and become distracted during an online course. Specifically, both unimodal models (using individual biometric variables) and multimodal models (combining multiple variables) were developed.

After comparing several algorithms, the results indicate that unimodal models relying on a single EEG signal or heart rate performed significantly worse than the model that integrated all EEG signals and heart rate. The second-best model used head pose data, suggesting that movement patterns provide valuable information about phone usage. However, the highest-performing model incorporated all multimodal data (EEG signals, heart rate, and head pose), achieving an accuracy of 91\%. This supports the idea that using multiple biosensors to capture different biometric signals leads to more accurate prediction models. These findings align with previous MMLA research, which highlights the value of multimodality \cite{Fierrez-Aguilar2005_ScoreNormalization,2023_SNCS_Human-Centric_Pena} in improving the performance of prediction models on various learning tasks \cite{giannakos2019multimodal,spikol2018supervised}.

In future work, we will focus on other distractions, such as browsing or checking emails, and we will aim to develop models that can detect a wider variety of learning events.
In addition, we will investigate other indicators that have proven valuable for detecting learner behaviors, such as eyeblink \cite{daza2024mebal2}, pupil dilation \cite{krejtz2018eye}, eye gaze patterns \cite{navarro2024vaad}, typing \cite{2022_TBIOM_TypeNet_Acien,2025_PR_KVC}, mouse \cite{2022_PR_BeCAPTCHA-Mouse_Acien} and touchscreen dynamics \cite{2018_TIFS_Swipe_Fierrez,2024_ESWA_swipeformer_Paula}, among others.
Furthermore, we will examine the integration of both local and global features \cite{daza2024deepface} and will attempt to extend this research to face-to-face learning with multiple students in order to develop more general models.

Finally, we plan to explore real-time interventions that notify learners when they are becoming distracted, helping them stay focused, and to evaluate the impact of such interventions on students' learning outcomes. In addition, we will evaluate the effectiveness of these interventions by examining their impact on learners' learning outcomes, including engagement, comprehension, and overall performance. Through this research, we hope to gain a deeper understanding of how personalized feedback can enhance the learning experience in online courses.

Furthermore, the methods developed here are useful in estimating distractions that arise from the undesirable use of smartphones in any monitorable task that requires concentration, not only computer-based online learning. In our future work, we will also explore how our methods apply to areas such as autonomous driving (in which the driver must pay full attention to the road sometimes \cite{2019_IBPRIA_HRdriver_JHO}) or traditional book-and-paper study.

\textbf{Acknowledgments.} Support by projects: Cátedra ENIA UAM-VERIDAS en IA Responsable (NextGenerationEU PRTR TSI-100927-2023-2), HumanCAIC (TED2021-131787B-I00 MICINN) and SNOLA (RED2022-134284-T).
%
%
%
\bibliographystyle{splncs04}
\bibliography{refs}

\end{document}